\documentclass{llncs}
\usepackage[utf8]{inputenc}

\usepackage{colortbl}
\usepackage{caption}
\usepackage{enumitem}
\usepackage{booktabs}
\usepackage{multirow}
\usepackage{algorithm}
\usepackage[noend]{algpseudocode}
\usepackage{hyperref}
\usepackage{graphicx}

\begin{document}

\pagestyle{plain}

\author{Francesco Ciclosi\inst{1}\orcidID{0000-0002-0191-888X} \and Silvia Vidor\inst{1}\orcidID{0000-0003-1838-8077} \and Fabio Massacci\inst{1,2}\orcidID{0000-0002-1091-8486}}

 \institute{University of Trento
\email{name.surname@unitn.it}
\and
Vrije Universiteit Amsterdam
\email{f.massacci@vu.nl}}

\title{Building cross-language corpora for human understanding of privacy policies}
\maketitle

\begin{abstract}
Making sure that users understand privacy policies that impact them is a key challenge for a real GDPR deployment. 
Research studies are mostly carried in English, but in Europe and elsewhere, users speak a language that is not English. 
Replicating studies in different languages requires the availability of comparable cross-language privacy policies corpora. This work provides a methodology for building comparable cross-language in a national language and a reference study language. 
We provide an application example of our methodology comparing English and Italian extending the corpus of one of the first studies about users understanding of technical terms in privacy policies. We also investigate other open issues that can make replication harder.

\end{abstract}

\section{Introduction} \label{introduction}

A well-developed literature exists in relation to the 
analysis of privacy policies (e.g., \cite{wilson,zaeem,zimmeck}), particularly concerning the development of tools to improve the writing \cite{hosseini} and clarity of the policies themselves \cite{fabian}. 

Specific studies on users’ understanding of privacy policies are instead less frequent (e.g.\cite{leicht,reidenberg,turow}). While the assumption that existing privacy policies are excessively long and complex is at the foundation of the majority of works on the topic \cite{amos,robillard,zaeem,zimmeck2}, the details of users’ misunderstanding of policies are less studied (e.g. \cite{ExampleStudy,vail}). For example, Tang et al. \cite{ExampleStudy} are the first to focus specifically on misconceptions related to technical terms used in privacy policies. 

Yet, the ability of citizens to understand what they accept (e.g., `are they waving their rights without even understanding it?') is a critical issue to gauge the actual success of privacy legislation initiatives such as the European Union General Data Protection Regulation (GDPR) \cite{GDPR}. The introduction of the GDPR in the EU has increased organizations' focus on complying with data protection principles. This privacy-driven approach progressively has grown the complexity of cybersecurity implementation and consequently their costs \cite{layton2019social}, encouraging the development of GDPR's compliance tools \cite{chatzipoulidis2019readiness}, and introducing security workers such as the Data Protection Officer \cite{9998077} (already existing in the past, but now well-defined and mandatory in all the EU member states). Because an organization must write and comply with privacy policies, it is crucial to assess whether the alleged beneficiaries of the protection granted by these policies (i.e., citizens) can actually understand what they are protected from (and what they are \emph{not} protected from).

A critical issue in this respect is that most studies use English and are based on the U.S. reality (e.g. \cite{sarne,ExampleStudy,wilson,zeadally}), which has a different culture as well as differently constructed legislation.

European Union policies are typically written in the national language of a country, and any study that is based on the English language would not accurately reflect the proper understanding of the users: we would be probing at the same time their understanding of English \emph{and} their understanding of the privacy issues. Still, using English as the reference language has advantages, particularly in ensuring that researchers from different countries can access and compare the work.

Given these premises, our main goal is to provide a methodology on how to generate comparable privacy corpora when dealing, on the one hand, with the English language and, on the other hand, with a National language (in this case, Italian).

In this work, we describe how we created such corpora and how to quantify the diversity within the corpora as well as a number of other open issues that makes replication studies way harder than one can initially think of.

\section{Related Work} \label{litreview}

Privacy policies have long been the subject of detailed scientific studies; with the advent of the Internet, online privacy policies have proliferated \cite{cecere,zeadally} – often in the form of long, complex and, easily misunderstandable statements – in tandem with the necessity to explain users’ data treatment for each online service. This, situation has led, to the necessity (that under some legislation, as in the EU, also became an obligation) to identify and develop methods to write privacy policies more easily, but also to make them more accessible to users of all backgrounds. Within the literature on privacy policies, two major issues are at the forefront of the discussion: the construction of adequate corpora for future analyses, and the development of automated tools for the analysis of existing policies and the drafting of future policies on the basis of a continuously evolving legal environment.

\paragraph{Privacy Policies' Corpora Selection.}
One of the main issues identified by the literature in proceeding with the (automated) analysis of existing privacy policies for their overall improvement is the lack of appropriate datasets from which to start such analysis \cite{amos,sarne}. Specifically, the problem for practitioners lies in the selection of policies that represent adequately the great variety in length, complexity and service coverage present among online privacy policies. The situation is further complicated by the presence of different legal backgrounds concerning privacy, for example between the European Union (where the content of privacy policies is mainly determined by the GDPR \cite{GDPR}) and China (where the Personal Information Protection Law (PIPL) sets similar requirements \cite{PIPL}) or the United States (where applicable laws vary between different States or circumstances). These differences can result in different policy structures and contents. As a consequence, the privacy policy of the same company can vary significantly depending on the country from which it is read. An additional factor of variety is the dimension of the company which the privacy policy is referred to, an element which generally impacts both the length of the policy and the frequency with which it is updated \cite{wilson}.

The methods used for policy selection vary between different works. Two main approaches emerge from the literature: the identification of criteria for the manual selection of representative policies (e.g. \cite{wilson}); and the development of web crawlers to extract the highest possible number of policies available online (e.g. \cite{amos}). The two approaches seem to reflect the distinct priorities of different studies conducted on privacy policies, which tend to focus either on the characteristics of the selected policies (a preference for ``quality'') or on the sheer amount of considered policies (a preference for ``quantity''). The second approach seems to be the most common in the literature, though a combined method -- establishing a set of quality criteria, often starting from a manual selection and analysis of a few policies generally selected on the basis of popularity, and using such criteria as a basis of action for a crawler -- is also employed often.

\paragraph{Tools for Analysis.}
A thriving strand of literature is then dedicated to automated tools aimed at, among others, analysing, creating, synthesizing, verifying the compliance or extracting specific elements of interest from privacy policies. The automated analysis of privacy policies has become a necessity both for the organizations writing the policies and for supervising authorities, but also for users requiring new manners to identify and understand the highlights of such policies \cite{delalamo}. Situated at the intersection between legal and technical domains, such analysis has recently turned to machine learning and text mining in order to automate and potentially improve a process that still relies heavily on human contribution \cite{sarne}.

Automatic privacy policy analysis is generally performed through natural language processing (NLP) techniques, which remain the dominant approach in the area. As reported by Del Alamo et al. \cite{delalamo}, NLP techniques can be divided into symbolic (or classic) and statistical (or empirical). The first starts from human-developed rules to process the policy's text and model natural language; the second, instead, applies mathematical techniques to previously-created corpora of policies to develop generalized linguistic models. The attention of practitioners is mostly directed to the statistical approach and to the creation of tools enabling its execution in an automatic manner \cite{zimmeck,zimmeck2}.

\paragraph{Readability and Users' Misconceptions.}
The academic interest over privacy policies is not limited to the manner in which they can be composed. In fact, the efficacy of a privacy policy does not depend only on its adherence to the legal requirements established by the countries of reference -- though that is indeed an important and necessary element -- but also on how understandable it is by the final recipient of the policy: the concerned website's user.\\ The capability of users to understand the content of privacy policies is analyzable in more than one way. The majority of the literature focuses on the measure of so-called ``readability'' \cite{ermakova,krumay}, which can be calculated according to a series of mathematical formulas or characteristics such as length, language complexity and univocal meaning.

Though definitely less studied throughout the years, a significant portion of scholarly investigation related to privacy policies has concerned users' misconceptions about the policies' meaning and function \cite{reidenberg,turow}. 

\section{A Reference Corpus in English} \label{originalstudy}
To start a reference corpus, we used the study by Tang et al. \cite{ExampleStudy}, who tried to investigate the understandability of privacy policies from the users' perspective, focusing in particular on specific technical terms commonly used in data use policies in the context of the USA. 

To achieve such an aim, the authors of \cite{ExampleStudy} ran three different studies:
\begin{enumerate}
    \item A qualitative pilot study to identify commonly misunderstood technical terms;
    \item A large-scale main study to test the respondents' understanding of the selected terms and their comfort with some data use practices;
    \item A small-scale follow-up study to support the main study's results.
\end{enumerate}

To select the 22 terms to be included in the main study, the authors of \cite{ExampleStudy} created a preliminary list of 57 terms obtained from manual analysis of the Alexa top 10 U.S. websites as of June 2020 (a common practice in studies on privacy policies \cite{reidenberg}) and some selected apps from the Android Play Store. The list was then validated through an automated analysis of a 3609 English-language policies corpus to verify the frequency of use of the selected terms. The 57 terms were also divided into 11 categories based on the macro-area they belonged to (e.g., crypto, storage, tracking). From the original study’s authors, we have obtained a spreadsheet including the technical terms (58 since one of them was missing from the original appendix) and the categories used in the pilot study. We also received the authors' original notes, where the policies in which they found the technical terms, the contexts of use, and a link to (the current version of) the relevant privacy policies are specified.

In the study by Tang et al. \cite{ExampleStudy}, the 57 terms, shown randomly based on their category, were included as part of the pilot study. In the pilot study, respondents from Amazon Mechanical Turk were asked to define the term they were shown. Based on the pilot results, 20 technical terms that were misunderstood the most (of which ten belong to the set of high-frequency terms commonly misdefined, while the others belong to the set of terms that the participants significantly misunderstood) were selected for the main study, together with two mostly well-understood terms. The main study was divided into two sections, asking respondents to answer multiple-choice questions defining some of the 22 terms and rate their comfort with some data use practices on a five-point Likert scale. The ratio of misunderstood to understood technical terms used for the survey is highly unbalanced (20-to-2), which may affect Tang et al. \cite{ExampleStudy} study's validity. From our perspective, we use them only as control elements to make the study comparable.

Due to the uncertainty connected to the definition of comfort, the follow-up study was then used to verify whether users' attitudes to policies changed using technical versus descriptive terms.

\section{Methodology} \label{methodology}
\begin{figure*}[!t]
\centering
\includegraphics[scale=0.9]{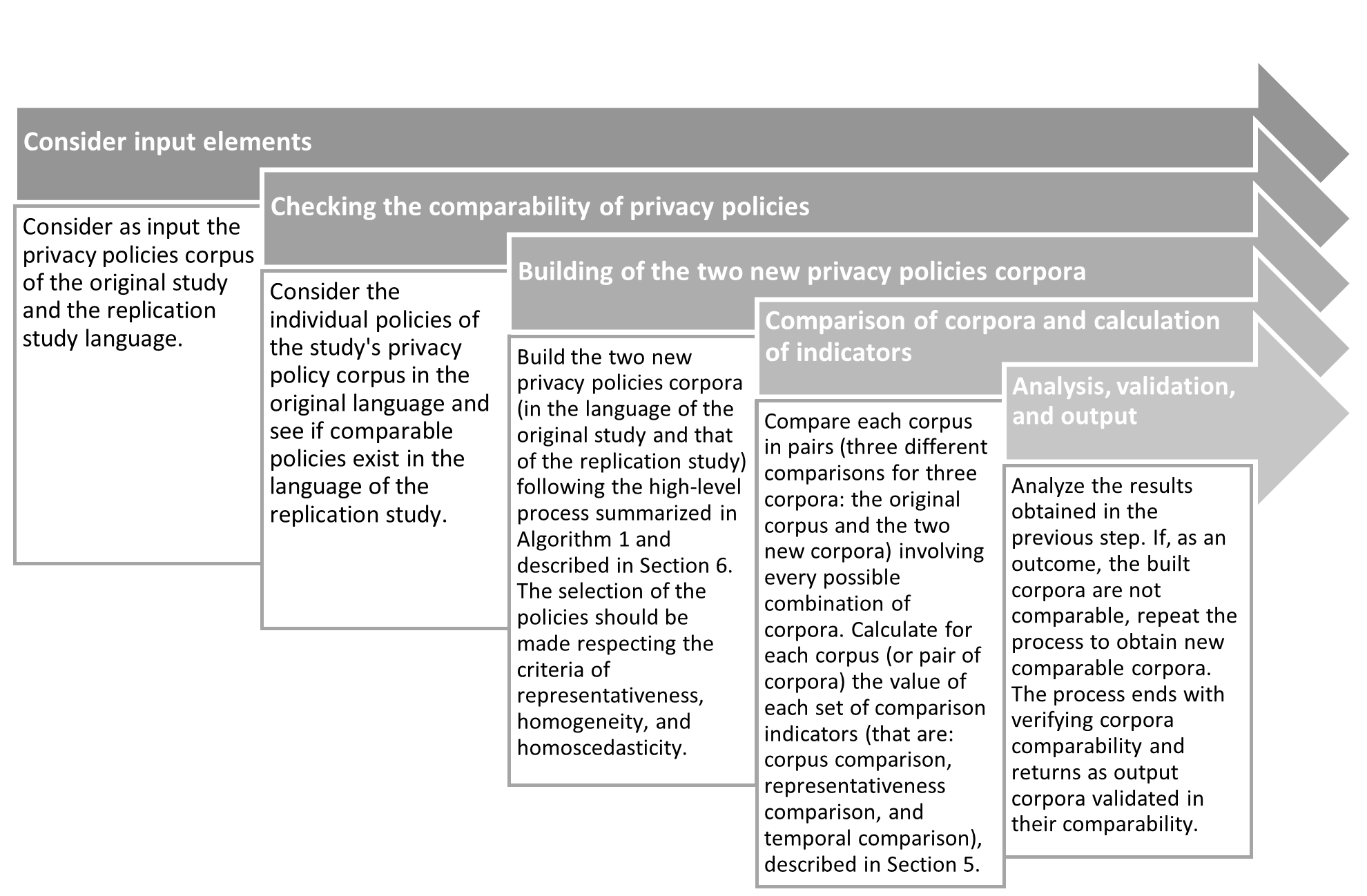}
\caption{The methodology to build comparable privacy policies’ cross-language corpora}
\label{fig:Methodology}
\end{figure*}

This section describes a usable methodology to build comparable privacy policies' cross-language corpora by mapping the policies of a corpus taken from a reference study into a new one that adopts the language of a future replication study. Figure \ref{fig:Methodology} represents this methodology's diagram summarizing inputs, outputs, main characteristics considered, and transformations steps involved. Indeed, a cross-national replication study needs to consider a new corpus of privacy policies comparable with the original one because it will probably not be available or existent in a hypothetical parallel corpus. Hence, in this case \cite{kohler2013statistical}, verifying the comparability of the two corpora is a precondition for using the new one.

The comparability check requires at least three different arguments. The first is the object to be compared, the second is another object to compare the first with, and the third is the respect used to compare these two objects. Hence, the third element is crucial, and its choice influences the comparison's result. Therefore, it is necessary to identify some properties concerning which they compare the corpora in comparing two corpora. 
There is the need to assess the comparability of a corpus with another that is not available since it will be composed of privacy policies used to replace some that are not available in the language of the future replication study. Therefore, before selecting policies based on the appropriate and specific criteria described below, it will be necessary to consider three data features \cite{kohler2013statistical}. These are representativeness, homogeneity, and homoscedasticity. The first feature relates to sharing a property between the two privacy policy corpora. The second feature requires that the data be homogeneous concerning the variables relevant to the purpose of the corpus. The third feature is related to the equal variance of the data across all classes.

Before a researcher can replicate a considered study, it is necessary to ensure that the corpora are comparable. 
Indeed, constructing multilingual corpora of privacy policies requires considering three different corpora: the \emph{original corpus} in the original study and in the original study's language, a \emph{source corpus} in the original study language and \emph{replication corpus} in the replication language.
Hence, it is necessary to make three different comparisons between three distinct corpora:
\begin{enumerate}
    \item between the original study's corpus and the new source corpus in the language of the original study;
    \item between the original study's corpus and the new replication corpus in the language of the replication study;
    \item between the new source corpus in the original study's language and the new replication corpus in the languages of the replication study.
\end{enumerate}

Identifying a new source corpus in the original language study is necessary because not all policies are usable/present in the replication language, and therefore, there might not be a correspondence between some/several policies in the original language and some policies in the replication language. As we illustrate here, this may happen because a company or institution in the original corpus does not exist in the replication country, or the company does exist but only have a policy in some languages and not the replication language. Only in very particular cases the original corpus and the new source corpus will coincide.

Usually, the availability of comparable corpora is more significant than parallel ones because they have the only requirement that their component documents cover related content in different languages \cite{li2013exploiting}. Many works in the literature have demonstrated the usability of comparable corpora in various areas; for example, in improving CLIR systems \cite{talvensaari2007creating} or in bilingual vocabulary extraction \cite{chebel2017bilingual}.
\cite{li2010improving} and \cite{skadicna2012analysis} have shown that an increase in the quality of a comparable corpus offers better performance to applications that refer to it. Moreover, \cite{li2013exploiting} highlighted the need to adopt methods that can qualitatively assess the quality of a corpus. This method allows moving beyond a naive approach based only on reasoning, which can lead to an a priori unpredictable level of performance.

Because of the problems highlighted in comparing different corpora \cite{kohler2013statistical}, we make use of solutions for measuring the distance between comparable corpora across languages derived from the one described in \cite{paramita2013methods}.

In our instance, we built new corpora of privacy policies (in Italian and English) from the privacy policies considered in the study taken as an example. These policies are accessible from the sites of the organizations that make them available through an interlanguage link (i.e., a link that relates documents on the same topic but written in different languages). This reflection allows us to extend to our case the considerations made in \cite{paramita2013methods} regarding retrieving comparable Wikipedia documents. Furthermore, a manual evaluation of the privacy policies retrieved and included in our new corpora (source and replication) showed that, in many cases, the privacy policies in the language of replication were created from a simple translation of the analogous source policies written in English. Indeed, as is evident from Table \ref{tab:policiescorpus}, most companies come from states whose native language is English.

Note that in our case, because the privacy policies are retrieved from the same website (simply choosing the correct language) or from different regional websites of the same company, it is probable that the considered corpora are parallel. However, this is not sure because, due to different legislation (from U.S. and EU), it is not taken for granted that a policy text is the translation of another policy text. Hence, we must assume the only certain fact that, in this situation, the two individually considered corpora are similar in some aspects; that is, they are comparable.

In our case, as suggested in \cite{paramita2013methods}, we focused on finding policies with a similar length (difference in word count less than 20\%) to increase the probability that they are comparable; that is, they are similar in structure and content.
Although most of the policies found have a similar length, there are some exceptions where this condition is invalid. Examples are in the policies of Yahoo, Amazon (IT and U.S., while the state is verified between the IT and NL-EN versions), and the USPS-Poste and Teamsystem-Force pairs (of which there is only one language version of the policies).

A replication gap between the referred example study and the replication study will always exist because some time has passed. This gap has to be manageable and acceptable. 
To this extent, we identified three groups of indicators to quantify such a gap and which we discuss in more detail in the next section. Then, we analyzed the outcomes provided by these three groups of indicators and
qualitatively discussed the real comparability of the corpora.

Finally, having verified the corpora's 
comparability, we focused on mapping the policies' terms. This activity was made manually and revealed 
some criticalities described in Section \ref{termsoverlapping} which is the ground for future work.

\section{Comparison Indicators} \label{indicators}
Based on the discussion in \cite{paramita2013methods}, \cite{kohler2013statistical}, and \cite{li2013exploiting}, we propose to use three groups of indicators: \emph{corpus comparison}, \emph{representativeness comparison}, and \emph{temporal comparison}, to qualitatively measure the privacy policies' corpora comparability.

The first indicators' group (\emph{corpus comparison}) measures whether the policies of the original study corpus are comparable with the corresponding (if they exist) policies in the cross-national study language.
To compare corpora, we are mainly interested in quantifying how many policies are comparable, replacement (or complementary), and destructured.
The indicator \emph{Number of Comparable Policies} measures the number of privacy policies referred to the same company for which there is both a text in the original and replication language. These texts must have a similar (comparable) structure.

The \emph{Number of Replacement Policies} indicator measures the number of policies in which the company from which the original study's policy is extracted does not exist in the replication country language. In contrast, the \emph{Number of Complementary Policies} indicator measures the number of policies for which does not exist a localized version of the policy the original study considered.
In the first case, if a product or service of a comparable company exists from the same industrial sector, the policy can be replaced with a new policy of these comparable companies. In the second case, if a localized privacy policy text of a \emph{comparable} company's product or services exists, this policy will be used. 

A privacy policy is considered complementary when it is added to a corpus to reflect any updates to a website rankings list (e.g., Alexa TOP10 U.S. or IT) that have occurred since the date of the original study.

The \emph{Number of Destructured Policies} indicator estimates how many policies where a policy in the target language exists; still, this text is not immediately comparable with the text in the original study language due to a significantly different structure. The more destructured policies, the harder any replication study is. To ensure better replicability across languages of a study, the number of comparable policies must be as high as possible.

The second set of indicators measures the representativeness of the corpora. It provides qualitative criteria to follow in determining how to replace policies in the original corpus that are not usable in the replication study corpora (source corpus and replication corpus). 

To measure the corpora's representativeness, we propose to do first a qualitative comparison using a table to compare the rank sources for each corpus. Indeed, it is crucial that the process used to build the new corpora be the same. Analogously, the rank sources used by the different corpora must be distinct from each other (because they are based on the constituent policies' language) and homogeneous (they must use the same source as, for example, Alexa Top10).
One possible replacement policy criterion is to investigate if a TOP web ranking used in a study carried out in a particular language also exists in a different language, such that both policies talk about the same terms.

For example, it is possible to consider the Alexa TOP10 website list in both languages. Another example is considering the ranking of top companies in a specific industry (e.g., the top banks) for each state to which the language used in the studies refers.

Finally, to consider the replicability of the original study, it needs to assess the temporal comparability of the corpora (third indicators' group). We evaluated qualitatively three indicators to measure this comparability: the \emph{Temporal Internal Consistency}, the \emph{Temporal Replication Gap}, and the \emph{Qualitative Replication Gap}.

We used the \emph{Temporal Internal Consistency} indicator to ensure that the source and replication corpora policies are all in the same narrow interval. This indicator measures within the same body of privacy policies the number of months between the publication date of the most recently updated policy and the one of the least recently updated policy.
The \emph{Temporal Replication Gap} indicator aims to highlight if the policy has changed. It measures the average number of months of the update time of policies in the source and replication corpora vs. the ones in the original corpus. This value is calculated only for the comparable or destructured policies because it has no sense for the other ones (complementary or replacement). Indeed, these policies are added contextually to the new source and replication corpora and are absent from the original one. Since the study may give different results because time has passed and many things have happened (e.g., changes in laws), the \emph{Qualitative Replication Gap} indicator is used to mark if major events happened from the original study and its replication.

\section{Reconstruction of policy corpus} \label{policycorpusreconstruction}
In the initial phase, we focused on looking for the correspondence between the Italian and English policies of the new cross-national source and replication corpora and the English policies of the original corpus we took from the example study of Tang. et al.\cite{ExampleStudy}. 
Hence, excluding the case in which the policies are comparable, we identified three cases that require attention. 
\begin{enumerate}
    \item Missing policy: there is no Italian policy; however, there is a corresponding Italian app or website.
    \item Missing company: there is no corresponding Italian app or website.
    \item Destructured policy: there is a corresponding Italian version of the original privacy policy, but we found that this has an entirely different structure.
\end{enumerate}

Henceforth, a team of two researchers analyzed the original corpus of policies and found correspondence rules to build comparable replication and source corpora of Italian and English ones. After defining our policy replication corpus for the Italian study, we concentrated on analyzing whether its policies were structurally similar to those in English of the original corpus. The high-level process to build the new replication and source corpora is presented in Algorithm ~\ref{alg:italianpolicylooking}. 
All the operations listed in the algorithm are done manually by humans.
When in Algorithm ~\ref{alg:italianpolicylooking}, the condition \emph{ItalianCompanyMissing} is verified, there is a case of a replacement policy. Analogously, if the condition \emph{ItalianPolicyMissing} is satisfied, there is no Italian policy, but there is a corresponding Italian app or website. Hence, this is a case of a complementary policy. Instead, if the condition \emph{ItalianPolicyDestructured} is verified, an Italian policy exists for a specific company. However, it has an entirely different structure from the original English policy (case of destructured policy).

If we need to substitute the English company with another of the same type (for example, another bank will replace a bank), we choose one that operates globally. Analogously, in case we need to substitute the missing privacy policy, we look for another related to a comparable company for the business sector, market segment, and type of product or service. With these choices, we have favored companies operating in the USA and Italy.

\begin{algorithm}
\caption{Look for the corresponding Italian privacy policy entry} 
\label{alg:italianpolicylooking}
\begin{algorithmic}[1]
\scriptsize 
\Procedure{Italian policy check}{$EnglishPolicyURL$}
    \State Retrieve English privacy policy text from $EnglishPolicyURL$
    \State Read English privacy policy text 
    \State Retrieve English company from $EnglishPolicyURL$
    \State Read English company 
    \State Search for the correspondent Italian company
    \If{$Italian Company Missing = True$}
        \State Choose a new company with the same type as the original company and a privacy policy in Italian and English
        \State Retrieve URL and text of Italian policy of new company
        \State Retrieve URL and text of English policy of new company
        \State Add new company's English policy to new English corpus as replacement
        \State Add new company's Italian policy to Italian corpus as replacement
    \Else
        \State Search for the corresponding Italian privacy policy of the original Company
        \If{$Italian Policy Missing = True$}
            \State Choose another privacy policy in Italian related to a company comparable to the original company
            \State Retrieve the URL and the text of the Italian policy of the new comparable company
            \State Add the original company's English policy to the new English corpus
            \State Add the new comparable company's Italian policy to the Italian corpus          
        \Else
                \State Retrieve the URL and the text of the original company's Italian policy
                \State Add the original company's English policy to the new English corpus
                \State Add the original company's Italian policy to the Italian corpus
                \State Analyze the Italian privacy policy structure
                \State Compare the Italian policy structure with the English one
                \If{$Italian Policy Destructured = True$}
                    \State Operate a manual measures activity of the privacy policy
                \Else
                    \State Classify the Italian privacy policy as comparable
                \EndIf
        \EndIf
    \EndIf
\EndProcedure
\end{algorithmic}
\end{algorithm}

We performed a manual analysis of the Alexa top 10 Italy websites as of November 2021, and analogously, we analyzed selected apps that, in the same period, had ranked better in the "most profitable games" category of the Play Store for Italy. After that, we compared these lists with the analogous ones (that refer to June 2020) in the original study for the U.S. privacy policies. To overcome the 17-month time gap between the establishment of the original corpus and the replication corpus, we made a further comparison by considering the content of the Alexa Top 10 U.S. ranking in the language of the original study as of November 2021.

From the original study, we considered 58 URLs obtained from 17 different companies' privacy policies. After this review phase, we removed six companies' privacy policies; however, we added twelve. This addition allowed us to improve the size of our corpus of policies, especially its specialization in the Italian reality and its focus on EU GDPR \cite{GDPR} concepts. We summarized these activities in Table \ref{tab:policiescorpus}. 
Our process in detail was as follows
\begin{enumerate} [label=\alph*)]
    \item the privacy policies of the apps \emph{Infinite Word Search} and \emph{Woody Puzzle} have been replaced by those of the apps \emph{Coin Master} and \emph{Empires \& Puzzles} that, in addition to having an Italian privacy policy, respectively (on November 23, 2021) ranked first and fourth in the "most profitable games" category of the Play Store for Italy;
    \item the privacy policy of the app \emph{Signal} is replaced respectively by the policy and by the FAQ documentation of the apps \emph{Whatsapp}  and \emph{Telegram}. The \emph{Signal} app has an Italian version but not an Italian privacy policy, so we have chosen to replace it with other messaging apps with similar characteristics, such as support for end-to-end encryption. Moreover, because for the app \emph{Telegram} there is not an Italian privacy policy but only an Italian FAQ documentation section in which the considered technical terms are present, we referred to this documentation;
    \item the privacy policy of the website \emph{Bank of America} is replaced by the \emph{Unicredit} one, that is a company of the same category and operating worldwide;
    \item the privacy policies of the \emph{Reddit} and \emph{Verizon} websites are replaced by the \emph{Vodafone} one, that is a company that provides analogous services;
    \item starting from one of the top 10 Alexa U.S. websites on November 2021, we added both the privacy policy of the related company and the one connected to a comparable Italian company (in particular the \emph{Force} U.S. company and the \emph{Teamsystem} Italian one);
    \item starting from one of the top 10 Alexa Italian websites on November 2021, we added both the privacy policy of the related company and the one connected to a U.S. comparable company (in particular the \emph{Poste} Italian company and the \emph{USPS} U.S. one);
    \item we added the privacy policies of the companies \emph{Zoom} and \emph{Microsoft} which have a website that appears both in Alexa top 10 Italian and in Alexa top 10 U.S. 
We chose policies from websites that appeared in the Alexa top 10 rankings for IT and U.S. to maintain equivalence between the original corpus, the new source corpus of English policies, and the new replication corpus of Italian policies. The original study's authors built their English technical terms corpus by manually analyzing the Alexa top 10 U.S. websites.
\end{enumerate}

\def\tick{$\surd$}

\begin{table*}
\centering
\caption{Composition of privacy policies corpus}\label{tab:policiescorpus}
\begin{minipage}{\textwidth}\scriptsize
This table describes the compositions of the original privacy policies corpus and the new English and Italian source and replication corpora. Columns named \emph{Orig. Tang et al.}, \emph{New US}, and \emph{New IT} refer to different privacy policies corpora. The first entry refers to the English-language privacy policies corpus analyzed in the original Tang. et al. \cite{ExampleStudy}study, while the second and third entries refer to the new privacy policies source and replication corpora in English or Italian, respectively. The \emph{Referred company} field shows the company (and eventually their product) whose privacy policy we considered. Finally, the \emph{Notes} field summarizes the rationale behind our choices in selecting the privacy policies.
\end{minipage}
\scriptsize
\begin{tabular}{p{4.1cm}p{0.8cm}p{0.5cm}p{0.5cm}p{4.5cm}}

\toprule
\textbf{Referred company} & \textbf{Orig. Tang et al.} & \textbf{New US} & \textbf{New IT} & \textbf{Notes} \\
\midrule
Yahoo & \tick & \tick & \tick & \multirow{11}{4.5cm}{We maintained it from initial study.}\\
King Games (app Candy Crush) & \tick & \tick & \tick & \\
LinkedIn & \tick & \tick & \tick & \\
Amazon & \tick & \tick & \tick & \\
Google & \tick & \tick & \tick & \\
Apple & \tick & \tick & \tick & \\
Facebook & \tick & \tick & \tick & \\
Wikipedia & \tick & \tick & \tick & \\
Twitter & \tick & \tick & \tick & \\
Ebay & \tick & \tick & \tick & \\
Firefox (documentation) & \tick & \tick & \tick & \\
\hline
Random Logic Games (app Infinite Word Search) & \tick & & & \multirow{2}{4.5cm}{We substituted policies because they do not exist in their Italian version.}\\
Athena FZE (app Woody Puzzle) & \tick & & & \\
Moon Active (app Coin Master) & & \tick & \tick & \multirow{2}{4.5cm}{Companies added from the PlayStore's rank on Nov. 2021.}\\
Zinga (app Empires \& Puzzles) & & \tick & \tick & \\
\hline
Signal & \tick & & & We substituted the policy because it does not exist in its Italian version.\\
Telegram (FAQ doc.) & & \tick & \tick & \multirow{2}{4.5cm}{Companies added to replace the \textit{Signal} one.}\\
WhatsApp & & \tick & \tick & \\
\hline
Bank of America & \tick & & & We replaced the policy because the respective Italian companies do not exist.\\
Unicredit & & \tick & \tick & Company added to replace the \textit{Bank of America} one.\\
\hline
Reddit & \tick & & & \multirow{2}{4.5cm}{We substituted companies with new ones focused on Italian reality.}\\
Verizon (cookie policy) & \tick & & & \\
Vodafone & & \tick & \tick & Company added to replace the \textit{Reddit} and \textit{Verizon} ones.\\
\hline
Zoom & & \tick & \tick & \multirow{2}{4.5cm}{Policies added from the top 10 Alexa Italian and U.S. on Nov. 2021.}\\
Microsoft & & \tick & \tick & \\
\hline
USPS (only US) & & \tick & \tick & It was added because it is a U.S. company with similar features and products to the Italian \textit{Poste}.\\
Poste (only IT) & & \tick & \tick & Company added from the top 10 Alexa Italian on Nov. 2021.\\
\hline
Teamsystem (only IT) & & \tick & \tick & It was added because it is an Italian company with similar features and products to the U.S. \textit{Force}.\\
Force (only US) & & \tick & \tick & Company added from the top 10 Alexa U.S. on Nov. 2021.\\
\hline

\bottomrule
\end{tabular}
\end{table*}

\section{Results and Comparison} \label{results}
In our example, we made three different comparisons between three different corpora to determine their comparability level.
The first corpus is the English-language corpus, the original used in the study by Tang et al. \cite{ExampleStudy}. The other two are cross-language corpora built (one, the source corpus, in English, and the other, the replication corpus, in Italian, which is the language of a potential replication study) from the first corpus. We aim to use them for a potential replication study on how humans understand privacy policies.

Comparing the English-language original corpus from Tang et al. \cite{ExampleStudy} with the source corpus built by us in the same language, we have 52,38\% comparable policies, 38,10\% complementary policies, 9,52\% replacement policies, and no destructured policies. The results are worst when comparing the original English-language corpus and the new replication one in Italian. In that case, we have 47,62\% comparable policies, 38,10\% complementary policies, 9,52\% replacement policies, and 4,76\% destructured policies. 

In contrast, comparing the two new source and replication corpora in Italian and English, we find these are well aligned. We have 85,72\% comparable policies, 9,52\% complementary policies, 4,76\% destructured policies, and no replacement policies. In all comparisons, the number of destructured policies is low, if not wholly absent.

Hence, rather than using Tang et al.'s \cite{ExampleStudy} privacy policies' source corpus for a cross-language comparison, we compare the two new source and replication corpora (built from the original one) that are more closely aligned.

Moreover, internal temporal consistency exists in the new source and replication corpora policies. Indeed, these policies are all in the same narrow interval because their release varied from June 2020 to September 2021. Furthermore, with few exceptions, the Italian and English versions of our policies were published simultaneously. The exceptions to this are Amazon (where the Italian version is December 2020 while the English version is February 2021), Facebook (with versions varying from August 2020 to January 2021), WhatsApp (varying from January 2021 to November 2021), or Vodafone (varying from July 2021 to November 2021).

For a correct calculation of the value of the "\emph{temporal replication gap}" indicator, it is essential to know the version and date of each privacy policy considered (both in the original corpus in the original study and those in the source and replication corpora in the replication study). 
Knowing the number of intermediate versions of privacy policies for each company is also helpful. In our example, there are many problems, and this indicator is not helpful. The original study \cite{ExampleStudy} did not specify the version and date of policy updates. We deduced the date of the considered policies by cross-referencing the updating date of the policy with the date of consultation of the TOP 10 of Alexa U.S. (June 2020). We know that this date is incorrect, but it still gives us an indicative idea.
Moreover, considering that not all companies expose the historical archive of policy versions, we can obtain a policy date to calculate the indicator in only 7 out of 11 cases. Other companies have made some intermediate versions of the privacy policy disappear from their website (for example, on the Facebook website, the previous versions are no longer available between September 9, 2016, and January 4, 2022). This fact generates another interesting problem because users have given their consent based on a no longer existing policy. We have not considered the case of Facebook because it presents an outlier value. Therefore, we used 6 out of 11 policies in calculating the indicator.
Interestingly, with only one exception (Google, for which there were four different versions in the face of a 17-month time gap), between the two studies, there was at most one version gap in the privacy policies considered. We calculated the temporal replication gap value only for the comparable or destructured policies because it has no sense for the other ones (complementary or replacement). Indeed, these policies are added contextually to the new source and replication corpora and are absent from the original one. The final value obtained for this indicator shows that the average number of months of the update time of the policy (comparable or destructured) is 16,5. This result is good because it shows that from the policies version available in the example study's original corpus and the new ones (source and replication corpora) built by us, there is only one version difference and a time lag between updates of just over a year.

Considering the qualitative replication gap indicator, it shows that in the time between the construction of the original corpus and of the pair source and replication corpora (June 2020 - November 2021), there have been no significant regulatory developments. For example, both the \textit{GDPR} and the \textit{California Consumer Privacy Act (CCPA)} predate the construction of the original corpus, while other laws, such as the \textit{Connecticut Data Privacy Act (CDPA)}, the \textit{Utah Consumer Privacy Act (UCPA)}, and the \textit{Virginia Consumer Data Protection Act (VCDPA)} was not yet in effect. Despite this, there have been some changes at the corporate level (e.g., the corporate change from \textit{Facebook} to \textit{Meta}) or at the policy level (e.g., additional information added regarding data retention in the Yahoo policy).

\section{Open Problems: Technical Terms May Not Match} \label{termsoverlapping}

After building a privacy corpus in a different language, the next step is replicating different studies. For example, the authors in \cite{ExampleStudy} arranged a survey based on 22 key terms derived from the pilot study's results to investigate the understanding of technical terms. 

We experienced many cases of technical terms that do not match and occur with extremely different frequencies.

The most egregious case are the English terms “\textit{personal information}” and “\textit{personally identifiable information (PII)}” (that were distinct in the considered study). Hence, we used a syntactic criterion both to be sure of their overlapping and overcoming it in a single Italian term, “\textit{dati personali}.”
In practice, we used a syntactic rule to identify the possible syntactic distinction of these terms. Hence, they are distinguishable if there is no correspondence between the two policies (English and Italian) or if there are orphans or widows. Otherwise, it is possible to conclude that they are the same thing and are codifiable with the same term in Italian.

To understand if the terms PII@US 
(\textit{personally identifiable information}) 
and PI@IT (\textit{informazioni personali},  \textit{personal information}) are 
equivalent, we manually looked for widows' or 
orphans' presence related to these technical 
terms.
An \emph{orphan} is a term that appeared in the Italian policy without a match in the corresponding clauses or sections of the English policy. A \emph{widow} is any occurrences in the English policy in which the term PII@US (or the term SPI - \textit{sensitive personal information} -) appears without the term PI@IT in the Italian policy. 

We also looked for implicit terms or pronouns that refer to technical terms. We found very few occurrences of those cases.

Other critical issues occur when technical terms do not appear or are rare in our corpora. For example, the term \textit{fingerprinting} does not appear in the Italian corpus and is useless for an Italian language survey. Instead, \textit{public information} and \textit{browser web storage} terms are rare in our corpora, but for an Italian survey, they are significant and can be used.

We illustrate these issues by considering the 22 terms from \cite{ExampleStudy}:
\begin {itemize}
\item two high-frequency terms that most users in the pilot study correctly defined;
\item ten high-frequency terms that the study's participants commonly misdefined;
\item ten additional terms for which the study's participants exhibited significant misunderstandings.
\end {itemize}
Using our new source and replication corpora of Italian and English policies, we investigate the differences between the frequency of occurrences in privacy policies of the most common Italian and English technical terms (the TOP 10 and the TOP 22 lists). More details are listed in Table \ref{tab:TopTermsInPolicies}.

\begin{table}
\centering
\caption{TOP 10 and TOP 22 lists of more frequent terms in the new Italian source corpus of privacy policies}\label{tab:TopTermsInPolicies}
\begin{minipage}{\textwidth}\footnotesize
This table describes the TOP10 and TOP22 lists of terms that appear most frequently in the new Italian source corpus of privacy policies. We used boldface to represent the terms belonging to the TOP 10 list. 
The order (\emph{Rank IT} column) of technical terms in the table follows the terms that appear most frequently in Italian language privacy policies. The table also shows the ranking of the most frequent technical terms in English language privacy policies (\emph{Rank U.S.} column). Corresponding to the Italian-language term (\emph{Term IT}) is the corresponding English-language term (\emph{Term US}). Finally, the table shows the cardinality with which we detected the term in the new replication and source corpora of Italian-language (\emph{\#Freq IT}) and English-language (\emph{\#Freq US}) privacy policies, respectively.

\end{minipage}
\footnotesize
\begin{tabular}{p{0.9cm}p{0.9cm}p{3.5cm}p{3.5cm}p{1.1cm}p{1.1cm}}
\toprule
\textbf{Rank IT} & \textbf{Rank US} & \textbf{Term IT} & \textbf{Term US } & \textbf{\#Freq IT} &\textbf{\#Freq US} \\

\midrule
1 & 1 & \textbf{Dati Personali} & \textbf{Personal information} & 1523 & 1430 \\
2 & 12 &\hfill of which  &  \textbf{Personally identifiable inform.} & 1012 & 79 \\
3 & 2 & \textbf{Cookie} / \textbf{Marcatori} & \textbf{Cookies} & 645 & 784 \\
4 & 4 & \textbf{Terzi} / \textbf{Terze parti} & \textbf{Third parties} & 618 & 360 \\
5 & 3 & \textbf{Informativa sul trattamento dei dati personali} & \textbf{Privacy policy} & 471 & 540 \\
6 & 14 & \textbf{Rettificare} & \textbf{Correct} & 170 & 60 \\
7 & 7 & \textbf{Indirizzo IP} & \textbf{IP address} & 136 & 153 \\
8 & 26 & \textbf{Disattivare} & \textbf{Deactivate} & 128 & 25 \\
9 & 9 & \textbf{Cifratura} / \textbf{Crittografia} & \textbf{Encryption} & 86 & 113 \\
10 & 5 & \textbf{Dati sull'account} & \textbf{Account information} & 60 & 324 \\
\hline
11 & 56 & Hash crittografico & Cryptographically hashed & 52 & 2 \\
12 & 36 & Transferimento di dati & Data transfer & 50 & 15 \\
13 & 10 & Affiliate & Affiliates & 48 & 87 \\
14 & 6 & Pubblicità mirata & Targeted ads & 47 & 157 \\
15 & 15 & Identificatori univoci & Unique identifiers & 45 & 54 \\
16 & 20 & API/SDK & API/SDK & 44 & 44 \\
17 & 16 & Dati relativi all'ubicazione & Location-related information & 42 & 50 \\
18 & 18 & Informazioni sui pagamenti & Payment information & 41 & 47 \\
19 & 40 & Resi anonimi & Anonymize & 40 & 13 \\
20 & 46 & Operazioni del dispositivo & Device operations & 39 & 6 \\
21 & 29 & Categorie particolari di dati personali / Dati sensibili & Sensitive personal information & 34 & 19 \\
22 & 19 & Dati di utilizzo & Usage data & 28 & 45 \\

\bottomrule
\end{tabular}
\end{table}

The correspondence between the TOP 10 English list and the TOP 10 Italian one is generally reasonable. 80\% of the terms included in the TOP 10 English list also belong to the TOP 22 Italian one (with a peak of 70\% of terms also included in the Italian TOP 10). Hence, only 20\% of terms in the TOP 10 English list do not belong to the TOP 22 Italian one. The correspondence between the TOP 22 English list and the TOP 22 Italian list is acceptable. 58,33\% of terms in the TOP22 English list belong to the Italian one, and 16,66\% of them also belong to the TOP10 Italian list. Hence, only 41,67\% of terms in the TOP22 English list are outside the TOP22 Italian one.
 
Out of 22 terms used in the Tang et al. \cite{ExampleStudy} example study's privacy policies English original corpus, ten of these (45,46\% of the total) appear in one of the TOP English lists of our new English source corpus (specifically, five in the TOP 10 and the other five in the TOP 22). While the remaining 12 terms (54,54\% of the total) are outside our source corpus TOP 22 English list. This outcome means that referring to our source corpus of privacy policies, more than half of the technical terms considered in the study taken as an example are rare. This result is crucial because our source and replication corpora are more focused on EU reality.

The situation is worse, considering the terms used in Italian policies. Only 7 (31,82\% of the total) of these terms appear in one of the TOP Italian lists (in particular, five in the TOP 10 and the other two in the TOP 22). The remaining 15 terms (68,18\% of the total) are outside the TOP 22 Italian list.

Hence, the technical terms used in the example study's survey by Tang et al. \cite{ExampleStudy} are rarely used in the privacy policies in Italian. This consideration is a crucial point that shows the need first to map technical terms and, secondly, to adapt the survey for replicating it to Italian-speaking respondents. These activities will be the subject of future research work.

\section{Conclusion and Future Works} \label{conclusion}
The first conclusion of our work is that a strict replication study in a different language is possible only up to a certain extent because some elements do not transfer from the original study to the new one. 

Therefore, our contribution is to discover and describe how elements (for example, the original corpus used in the survey) can be adapted from an original study to a new one to understand what is needed to maintain relative comparability of the components of the study, such as the privacy policy corpora.

Still, the results obtained in our case study confirm that by taking due care, it is possible to build cross-language source and replication corpora of privacy policies usable for research purposes, such as investigating how humans understand their content. 

The outcomes of this work will be the base for further research activities. Future interesting work is to investigate the validity of automated analysis approaches, i.e. whether just looking for term occurences in a policy would yield a correct analysis. Precision and recall values achievable in the automated coding of technical terms should be compared from those derived by manual coding.
A second aspect is mapping technical terms in different language corpora privacy policies. Our preliminary analysis has already shown that this is not so obvious. Future work could define some indicators to identify the irrelevant, infrequent, overlap, and unreliable technical terms that must be removed or replaced according to the case. This information will help, for example, to replicate the understandability questions from Tang et al. \cite{ExampleStudy} to Italian-speaking people who interact with Italian privacy policies.

\section*{Acknowledgement}

The authors would like to thank Eleanor Birrell and Ada Lerner for providing us their raw privacy corpus used in their paper \cite{ExampleStudy}. Without their time and expertise this paper would not have been possible. This research was partially supported by the European Union and the H2020 project  830929 (CyberSec4Europe).

\subsection*{Dataset Availability}
The datasets generated and analysed during the study will be made available on Zenodo upon acceptance.

\subsection*{CRediT statements}

\emph{Conceptualization:}	FC, FM, SV;
	\emph{Methodology:} FC, SV;
	\emph{Software:} SV, FC;
	\emph{Validation:} FC, SV, FM;
    \emph{Investigation:} FC, SV;
    \emph{Data Curation:} FC, SV;
    \emph{Writing - Original Draft:} FC, SV;
    \emph{Writing - Review \& Editing:} FC, SV, FM;
    \emph{Visualization:} FC;
    \emph{Supervision:} FM;
    \emph{Project administration:} FM, SV;
    \emph{Funding acquisition:} FM.

\bibliographystyle{abbrv}
\bibliography{sources}

\end{document}